\documentclass[10pt, amssymb,preprint,twocolumn]{revtex4}
\usepackage{hyperref}
\usepackage{amsmath}
\usepackage{graphicx}
\usepackage{amssymb} 
\usepackage{amsfonts}
\begin{document}

\setlength{\parskip}{0mm}

\title{Heuristic rule for constructing physics axiomatization\\
{\it \small Essay written for the FQXi contest on "What is Ultimately Possible in Physics?"}}
\author{Florin Moldoveanu}
\affiliation{Department of Theoretical Physics, National Institute for Physics and Nuclear Engineering, PO Box MG-6, Bucharest, Romania $^{\dag}$\footnotetext{$\dag$ On leave.}}
\email{fmoldove@gmail.com}

\begin{abstract}
Constructing the Theory of Everything (TOE) is an elusive goal of today's physics. G\"{o}del's incompleteness theorem seems to forbid physics axiomatization, a necessary part of the TOE. The purpose of this contribution is to show how physics axiomatization can be achieved guided by a new heuristic rule. This will open up new roads into constructing the ultimate theory of everything. Three physical principles will be identified from the heuristic rule and they in turn will generate uniqueness results of various technical strengths regarding space, time, non-relativistic and relativistic quantum mechanics, electroweak symmetry and the dimensionality of space-time. The hope is that the strong force and the Standard Model axiomatizations are not too far out. Quantum gravity and cosmology are harder problems and maybe new approaches are needed. However, complete physics axiomatization seems to be an achievable goal, no longer part of philosophical discussions, but subject to rigorous mathematical proofs.\end{abstract}

\maketitle
\section{Introduction} 

Mathematics is used extensively in physics, and the search for a deeper connection between them originated with David Hilbert who in 1900 proposed a set of problems at the International Congress of Mathematicians which would advance the discipline of mathematics \cite{Hilbert1}. Among them, problem six requested the axiomatization of physics. While axiomatizing various branches of mathematics was achieved relatively easily, physics axiomatization faces huge obstacles. First and foremost, after the Galilean revolution, physics is an experimental science. Second, by G\"{o}del's incompleteness theorem \cite{GodelTh}, mathematics is infinite and there is no universal axiomatization of mathematics. It is conceivable that physics is also infinite in the same sense and axiomatizing physics appears hopeless from the start. And third, mathematics deals with abstract relationships, while physics deals with the real world. How can one expect that a pure mathematical technique as axiomatization to be applicable on such a different domain?

Despite the difficulties above, the hope of axiomatizing physics and constructing the TOE in physics is alive mainly because mathematical structures have been very successful in describing nature and gauge theory emerged as the universal language of the Standard Model. Also the properties of elementary particles or the values of the Standard Model parameters are the same wherever we measure them, and while we do not fully understand them right now, there must be some rigorous mathematical reasons for their values.

\section{Existing heuristic methods}
Before attacking the problem head on, let us investigate what tools, if any, we have at our disposal in constructing a physical theory. One commonly used criterion for a new physical theory is mathematical beauty. Another is simplicity or Occam's razor. Unfortunately, while useful in certain cases, both of them do not \emph{always} work in practice as seen by simple counter-examples. Take mathematical beauty. Natural numbers have a very neat modern axiomatization. But consider doing large number multiplications using Roman numerals instead of modern Arabic numerals. Same mathematical structure, yet it is no wonder why Romans did not make significant mathematical contributions. Occam's razor does not fare any better in always distinguishing between valid and invalid physical theories. Newtonian mechanics is far simpler than quantum mechanics, and yet it is just an approximation that gives the completely wrong picture in terms of local realism and the nature of randomness. 

The problem of any heuristic method is that there is no universally valid approach toward discovering new theories. It would be nice if one can use a prescribed approach and generate all kinds of new theories automatically, or even better, program a computer to do it for us. But by G\"{o}del theorem, this is simply impossible. Still one can look at the lessons of the past, and attempt to draw conclusions from them. A typical starting point is to work on some problem that does not fit well within the established paradigm. The next step is to find some way of reasoning about the problem that would lead into the undiscovered theory. This step should not rely on clever tricks, or sudden inspiration, but should be a systematic research program and each step should follow easily from the previous one. Of course, this is easier said than done, and many such approaches are not very fruitful and would not lead to anything significant, but this is the nature of research.

So what would be the initial problem for constructing the theory of everything? The concrete starting problem is beginning to construct the physics axiomatization, and not obtaining a grand unification theory (GUT), or explaining everything from the beginning. If the axioms of physics are identified correctly, working out their consequences will eventually lead to the ultimate TOE if it indeed exists. But why is selecting the postulates of physics not an impossible hard problem? Starting from nothing it is, but we do have a critical mass of mathematical results which when put together point the way towards the solution. Those results are unfortunately not well known because they were \emph{all} obtained outside the mainstream of physics research. With the benefit of hindsight and ignoring the original justification of their research programs, here is how we can approach the problem of selecting the axioms of physics. Mathematics and physics have very different characteristics. The natural starting point towards physics axiomatization (a pure mathematical technique on a domain that is not math) is to investigate in what way physics is similar or different than math. 

\section{Comparing physics and math, and a new heuristic rule}

Comparing physics and mathematics is not a new problem, and some results are already available. Several authors have explored the idea that mathematics and physics are dual objects \cite{McCabe1},\cite{Heller1},\cite{Majid1},\cite{Tegmark1}. While certainly attractive, particularly the usage of category theory, Hopf algebras, and non-commutative geometry, since there is no universal mathematical axiomatization, this approach is not very useful for constructing the TOE, because to the extent that is right, it can only generate a no go theorem on physics axiomatization. 

Instead on trying to solve the nature of reality, a much more useful approach is not to seek the similarities, but the differences. And this is the main idea of this essay. Stated formally:

\vspace{5mm}\noindent{\bf Heuristic rule}. {\it Identify all mathematical properties of the physical world that are universally valid in the real world and are not universally valid in the abstract world of mathematics.}\vspace{5mm}

Typically, those mathematical properties are called physical principles, to recognize their elevated status from mere technical postulates. Ideally, all physics should be derivable only from them without the need for the additional scaffolding of technical postulates, but this is a work in progress. From several independent lines of research, three physical principles were clearly identified so far. Let us present them and their consequences. While their research is not new for the most part, combining them under the heuristic rule above creates a new cohesive paradigm on how to approach building the ultimate theory of everything that no longer relies on inspired guesses, but on a much more straightforward systematic approach.

\section{Physics principles and their consequences}

\subsection{Events and the universal truth property principle}

In general in mathematics, the truth value of a statement depends on the context. For example, the statement that two parallel lines never meet is true in Euclidean geometry, and false on Riemannian geometry. The mathematical meaning of truth is coded by the Tarski theorem \cite{Tarski1} which roughly states that inside an axiomatic system, one cannot define the truth value of its own predicates. Thus, in mathematics, truth means that something is derived from axioms, while in the physical world truth is usually defined as something corresponding to reality and has a ubiquitous non-trivial universal property. In physics, events occurring in the four dimensional event manifold are true for all observers and across all contexts. This universal truth property (or universal non contextuality) is a remarkable property enjoyed only by the real world. The universal truth property is clearly distinguishing between the real world, and the Platonic world of mathematics.

\subsection{Interaction and the unrestricted composability principle}

In the real world, any two physical systems can potentially interact with each other. In other words, there are no "island universes" (even black holes evaporate in the end and the mater trapped behind the event horizon can ultimately interact again with the rest of the universe). In the mathematical world on the other hand, there are sharp boundaries between different axiomatic systems. But interaction alone lacks concrete predictive mathematical power. We need to take the interaction observation a step further and note that all the physical laws that apply to any two physical systems, apply the same way to the total composed system. Take two quantum systems. Put them together and the resulting system is still a quantum system subject to the same Plank constant as the individual systems. This may look to be a trivial observation, but in effect it has far reaching consequences because it imposes major constraints on the potential physical laws allowed. In mathematical terms, it imposes the necessity of a tensor product which in turn severely restricts the choices of relevant mathematical structures that can describe the real world. For example (paraphrasing Grgin), two circles cannot be combined in any way such that the composite system is another circle. The same is true for all of the infinite number of mathematical structures with only a few "sporadic" exceptions.

\subsection{Unrestricted complexity and the deformability principle}

Within an axiomatic system in math, the amount of information is restricted by the original axioms. This is why computers for example cannot be programmed to be creative. In the real world on the other hand, there are unlimited ways in which matter distributions are allowed (as long as we stay above the Plank scale and the event manifold is continuous). Unrestricted deformability of matter distributions will impose constraints on the allowed geometrical structure of space-time.   

\subsection{Consequences of the three principles}

All of the principles above define sharp differences between the Platonic world of abstract mathematics and the physical world. It can be shown that they impose severe constraints on selecting which mathematical structures are useful to describe the real world. In this approach, uniqueness results follow from those principles, and we may now ask with mathematical precision questions like: "what is the origin of complex numbers in quantum mechanics?", "is quantum mechanics unique?", "why is the event manifold Riemannian and not Finslerian?", "what is the nature of time?", "why our universe has three spatial and one time dimension?", "what is the origin of electroweak symmetry?". All the above questions have clear mathematical answers. However, for now their answers involve additional technical assumptions, some with more additional assumptions than others. One goal of constructing a final theory of everything is to eliminate those additional assumptions as much as possible. Also the anthropic principle was considered in the past as a potential solution to the landscape problem of string theories \cite{Susskind1}, but this is highly controversial and will be discussed at the end of this essay. It is not clear at this point if the number of physics principles will be finite or there is an entire hierarchy of such principles to achieve a full theory of everything. From the initial outlook, since so much is explained by only three principles, it is very likely we only need a small finite number of principles to construct the theory of everything. The missing ingredient for now is how to approach the quantum gravity problem and the physics at the Plank scale. Let us now present the origin of the three principles. 

\subsubsection{The universal truth property}
The universal truth property was first investigated by the present author and started from trying to understand what role G\"{o}del incompleteness theorem can play in physics. From a high level overview, this theorem showed for the first time that truth and provability are separate concepts and there are true, but unprovable mathematical statements. Now one can easily equate in physics events with true statements, and causality with provability. Moreover, in a Minkowski space, all the events inside the causal cone are potentially accessible (provable), but the ones outside are not (unprovable but true) and in this high level sense, there is a deep connection between relativity and the incompleteness theorem. Investigating this connection led uniquely to the problem of time travel in general relativity and the nature of time. The foundation for G\"{o}del's proof was the Liar's paradox, or any self-referencing antinomy. A similar problem occurs in solutions of general relativity with closed timelike curves (CTC) and here one has the "grandfather paradox": a person goes back in time and prevents his grandfather to ever meet his grandmother thus preventing his own birth. The solution is simple: somehow forbid all paradoxes. This is usually called the Novikov principle \cite{Novikov}. From a pure classical point of view, there are two possible approaches: forbid initial conditions leading to paradoxes \cite{RamaSen}, or have "ghost solutions" \cite{Krasnikov1}, both unphysical. From the quantum point of view, the situation is subtler since one can have a superposition of both contradictory states, and the grandfather paradox does not apply directly. Instead in this case it manifests itself as a violation of unitarity. Since we still lack an ultimate theory of everything, we cannot conclude with certainty that time travel does not exist. But what we can do is to find consequences of the Novikov principle. In the classical case, this principle demands infinities \cite{Florin2} which in turn prohibits any quantum solutions compatible with time travel in the range of the validity of the correspondence principle. In other words, time travel may exist only at Plank scale, but the very notion of time is not defined there anyway.

The universal truth property can be shown to create a constraint on the event manifold that manifest itself as global hyperbolicity \cite{Florin1}, typically called time. Time is necessary to prevent the logical inconsistencies that characterize the Platonic world of math, and which do not exist in the real world. But wait a minute. How can we state that mathematics is full of contradictions? Is not math supposed to be all precise and exact and free of errors? Yes it is, but only within a given boundary of an axiomatic system. Different axiomatic systems can be and are at odds with each other. Here is how.  The incompleteness theorem shows that at least in some cases one can always find a new statement (or in the mathematical terminology a predicate) $p$, which cannot be proved or disproved within the existing axiomatic system. If the predicate is then added as a new axiom, the process can be repeated again in the extended axiomatic system. Since the new axiom can be added as either ($p$) or (not $p$), the process generates two new incompatible axiomatic systems. This is why mathematics in infinite and cannot be organized in a coherent system. The universal truth property on the other hand completely eliminates the objection against physics axiomatization from the G\"{o}del's result point of view because if in the physical world everything is consistent, then it might be axiomatized successfully. Moreover, the constraint it generates manifests itself as time. 

There is another matter to consider. Universal truth property does not show how time can exist, only that it must exist. There are strong indications pointing towards a quantum mechanical origin of time from relativistic quantum mechanics and the CPT theorem.

\subsubsection{Unrestricted composability principle}
This principle has its origin in the algebraic approach to quantum mechanics and was pioneered by Emile Grgin \cite{composabilityPaper}, \cite{GrginBook1}. Quantum and classical mechanics are using two products in general: one symmetric, and one anti-symmetric. For quantum mechanics the symmetric product is the Jordan product and it describes the algebra of observables, while the anti-symmetric product is the commutator and it describes the algebra of generators. Similarly for classical mechanics one has the regular function multiplication and the Poisson bracket. When two physical systems interact, the composed system should also be described in the same formalism, meaning it should also have a symmetric and anti-symmetric product. There are only three kinds of composability classes: elliptic (quantum mechanics), parabolic (classical mechanics), and hyperbolic (split-complex quantum mechanics). The hyperbolic quantum mechanics is unphysical because it violates von Neumann uniqueness theorem \cite{vonNewman}, and implicitly the universal truth property. Eliminating the classical case is a non-trivial problem and there are indications that this may ultimately be the case because classical general relativity suffers from major problems of singularities and time travel solutions. For now we only need to assume a technical separation principle between classical and quantum mechanics and any principle will do such as: the state space of quantum mechanics is continuous \cite{Hardy1}, or the maximum evidence accessible through experiment is not allowed to exceed
some finite upper bound \cite{Rau2}, or simply that quantum mechanics is a different composability class than classical mechanics. In fact, quantum mechanics (both non-relativistic and relativistic cases) can be completely obtained from three postulates \cite{Florin3}: composability, consistent reasoning about a physical system (a consequence of the universal truth property), and a technical separation axiom. The additional technical assumptions are linearity, reality of the spectra, the existence of the Jordan algebra, and the necessity of Lie algebra describing time evolution. 

The advantage of the algebraic approach is that it encompasses both the  non-relativistic and the relativistic cases. In the non-relativistic case complex numbers play the central role, along with the standard Hilbert state space. Following Hardy's analysis \cite{Hardy1}, if $N$ is the number of states, and $K$ is the associated number of degrees of freedom, from composability $K = N^r$ with $r$ a positive natural number. $r=1$ corresponds to classical mechanics and $r=2$ corresponds to non-relativistic quantum mechanics. Jochen Rau showed that $r >2$ demands sudden jumps in probabilities \cite{Rau2}. In fact those jumps correspond to creation and annihilation of particles in a second quantization approach of {\it relativistic} quantum mechanics. If in non-relativistic quantum mechanics one has the Born rule and complex numbers, in the relativistic case ($r=3$) they get replaced by the Zovko rule and quantions \cite{GrginBook2} - a non-division algebra. The $N^2$ dimensional Hilbert space gets replaced with an $N^3$ degrees of freedom non-commutative geometric state space$^{1}$\footnotetext[1]{$r$ is given in general by the valence of a generalized Born rule algebraic norm. In classical mechanics the state space is discrete and there is no cross talk between distinct states resulting in a Kroneker delta result. In the non-relativistic case, the result is the norm of a complex number, while in the relativistic case it is the norm of a quantion generating a current probability density. The divergenceless current probability relativistic case reduces itself to the standard Hilbert state space and gives rise directly to the Dirac equation and $SU(4)$ and very possibly to the Klein-Gordon equation and $SU(2,2)$. The general divergent current probability case is a much more difficult open problem.}. Loosely speaking, the Feynman diagrams are only $N^2$ "flatlander" Hilbert "slices" in an $N^3$ degrees of freedom state space. Mathematically this is more complicated because the quantions are not a commutative unital C* algebra and therefore the relativistic case lacks the Hausdorff's property and we are in the realm of non-commutative geometry. Discovered first in 1882 by Benjamin Pierce \cite{Pierceref}, quantions are related to Clifford algebras and were also named the "space time algebra" by David Hestenes \cite{Hestenes1} who studied them from the Clifford algebra point of view. The most remarkable property of this algebra is that it contains the electro-weak symmetry $SU(2)\times U(1)$ and seems to naturally explain the particle chirality properties and (for now) the semi-classical aspects of the electroweak interaction. Starting from composability Grgin was the first to prove the central unique role of this algebra to the structural unification of quantum mechanics and relativity, and that \emph{this unification is only possible in a four dimensional space-time}. This is remarkable especially considering the string theory results. In string theory the space-time dimensionality is a multiple of the Bott periodicity (8) plus 2 and the minimum dimensionality it can get is 10. Compactification can reduce this to the usual 4, but why is 4 special? Why not 5, or 6, etc? The answer lies in the uniqueness proof of quantions. 

Regarding the link between quantum mechanics and time, first Connes \cite{Connes2} studied the global and Alfsen, Hanche-Olsen, and Shultz \cite{Alfsen1} the local orientability of quantum mechanics related to a C* algebra. This is related in the relativistic case with a spin factor Jordan algebra and the Minkowski cone. In other words, unitarity and coherence demands global hyperbolicity of time-space. If not, in a CTC space a particle can be generated at a point, travel around a closed loop acquiring a phase difference and being re-absorbed at the place of generation breaking unitarity \cite{Boulware}. In the quantionic approach, the transition from the SO(2,4) space to the Minkowski SO(1,3) space is achieved via "Grgin complexification" incorporating directly the CPT discrete symmetry \cite{foundationPaper3}. If relativistic quantum mechanics demands the existence of time, then there should be also a link between the standard Hilbert space non-relativistic quantum mechanics and time. The work of Connes and Rovelli \cite{Connes1} clarifies this relationship. Using the Tomita-Takesaki theorem, they propose the emergence of time from the thermodynamical state of the system in a general covariant theoretical framework.

\subsubsection{Deformability principle}
General relativity is described using Riemannian geometry which singles out the standard quadratic metric. Why is nature not described by a Finslerian geometry, or any other kind of geometry? This question was answered by Rau \cite{Rau1} and to answer it he introduced the deformability principle. His analysis was based on dimensional analysis of Lie groups among other things and he proved that arbitrary mass distributions demands the $SO(p,q)$ orthogonal group. To single out the $SO(p, 1)$ group one needs the universal truth property principle, and to single out further the $SO(3,1)$ group one needs composability and the search for non-unitary realizations of quantum mechanics  via internal complexification in a $SO(2,4)$ space \cite{GrginBook1}. 

So far this principle has the largest number of additional technical assumptions and the main challenge is to see how it may hold at Plank scale. In passing we note that in quantum mechanics one may still encounter infinite complexity at Plank scale because measuring incompatible properties results in the prior information being erased.

\section{Discussions and Conclusions}

We have seen that starting from asking how physics is different than math, mathematical structures can be uniquely identified. Obtaining the necessity of space, time, non-relativistic and relativistic quantum mechanics, electroweak symmetry and the dimensionality of space-time are only the first steps. Additional results indicate that the electroweak force beyond the semi-classical aspects, strong force and the Standard Model axiomatizations are achievable in the near future. There are three basic problems that remain to be solved: prove that one cannot construct a coherent theory based on classical mechanics only, eliminate the additional technical postulates, and the hardest problem of all: solve the quantum gravity and cosmology problem.

For quantum gravity four basic major research approaches are currently considered, all with important open problems. First, the string approach (I include here also the twistor research) is the most developed, but it has major problems in terms of the minimum space-time dimensionality and finding a unique vacuum. Second, background independent approaches show promising results, but their major problems are obtaining the large scale continuum event manifold and predictions of Lorenz violations. Third, the non-commutative geometry offers a simple axiomatization of the Standard Model including gravity, but its recent prediction of the Higgs boson mass was ruled out by experiments, and the approach needs to be revisited. Last, the geometric algebra approach led to the development of the Gauge-Theory of Gravity, but this is still in its infacy having to prove it passes all of the experimental tests of general relativity.

One physical principle was proposed before to justify the $10^{500}$ potential solutions of string theory. The anthropic principle seems to explain this problem, but there is a major issue: it lacks predictive power. This principle can be also applied for example to the pre-quantum chromodynamics era in physics and justify the state of knowledge at that time. The landscape may have a true physical reality and the anthropic principle may be true, but it has no real value towards constructing physics axiomatization. 

The author's (biased) personal preference for the quantum gravity problem is to investigate more the algebraic approach to quantum mechanics and to find additional links between quantions and non-commutative geometry and between non-commutative and conformal geometry. If the strong force has also a quantionic origin, then the Pati-Salam GUT is the preferred GUT approach. However it seems that the real problem is not force unification going up the energy scale, but force separation going down the energy scale. The state space of the Big Bang and the center of a black hole should be described in the relativistic $N^3$ degrees of freedom space, and not into the standard $N^2$ dimensional Hilbert space.

The three axioms of physiscs identified so far have significant mathematical consequences and also philosophical consequences in term of understanding the nature of reality. Suppose for the sake of argument that physics and math are indeed dual objects, each one completely describing the other. From universal truth property, we can then understand reality as the Platonic world of math ordered by time to eliminate the self-referencing paradoxes. This ordering can be achieved in may ways, and composability is reducing the degeneracy of the ordering by demanding a coherent description of nature at all scales. This goes a long way towards proving the uniqueness of the real world. The simple finite dimensional mathematical objects compatible with those two principles as well as the rest of the infinite world of mathematical structures are prevented to play a universal role by the deformability principle, but they may still get to play a role in special cases. Of course this is just a nice philosophical speculation, but we should follow the math into proving stronger uniqueness results. While the new axiomatization paradigm changes the perspective, working out all its consequences in the ultimate hypothetical Theory of Everything is still faced with the same old problems. But now we can also ask well posed uniqueness questions as well as pursuing unified interaction theories. Additional principles may be identified by the heuristic rule, or by other future approaches. If this new line of research will turn out to be fruitful, the uniqueness questions should provide additional insights into solving old puzzeles. When experiments are no longer needed or possible, we will enter a post Gallilean era when physics would no longer rely exclusively on experiements as the ultimately judge of success. Math never depended on experiments and whenever physics is experiment independent, it should be done more like math: with increased rigour and safeguarding a diverse approach towards research directions.

\section{Acknowledgments} 
This essay has benefited greatly and was inspired in part by the research of Emile Grgin and in particular by the disclosure of his heuristic approach towards achieving the structural unification of quantum mechanics and relativity \cite{GrginBook1}.

\end{document}